# Sputtered 2D transition metal dichalcogenides; from growth to device applications

**Merve Acar[1] and Emre Gür[*2]**

[1]Department of Electrical and Electronics Engineering, Faculty of Engineering, Atatürk University, 25240, Erzurum, Turkey

[2] Department of Physics, Faculty of Science, Ataturk University, 25240, Erzurum, Turkey

## Abstract

Starting from graphene, 2D layered materials family has been recently set up more than 100 different materials with variety of different class of materials such as semiconductors, metals, semi-metals, superconductors. Among these materials, 2Dsemiconductors have found especial importance in the state of the art device applications compared to that of the current conventional devices such as (which material based for example Si based) field effect transistors (FETs) and photodetectors during the last two decades. This high potential in solid state devices is mostly revealed by the transition metal dichalcogenides (TMDCs) semiconductor materials such as $MoS_2$, $WS_2$, $MoSe_2$ and $WSe_2$. Therefore, many different methods and approaches have been developed to grow or obtain so far in order to make use them in solid state devices which is a great challenge in large area applications. Although there are intensively studied methods such as chemical vapor deposition (CVD), mechanical exfoliation, atomic layer deposition, it is sputtering getting attention day by day due to the simplicity of the growth method together with its reliability, large area growth possibility and repeatability. In this review article, we provide benefits and disadvantages of all the growth methods when growing TMDC materials, then focusing on the sputtering TMDC growth strategies performed. In addition, TMDCs for the FETs and photodetector devices grown by RFMS have been surveyed.

[*]**corresponding author e-mail address:** emregur@atauni.edu.tr
**Phone:** +90 442 231 4082
**Fax :** +90 442 231 4109

**Keywords;** Sputtering, 2D materials, transition metal di-chalcogenides, 2D field effect transistor

## 1. Introduction

A new era in science has opened up by discovery of of graphene in 2004 with the publication named “Electric field effect in atomically thin carbon films” by Novoselov *et al.* [1] which is one of the top 100 papers cited of all time [2]. This new topic has launched variety of different 2D materials within a few years after which it has dominated the number of publications and research performed throughout the last two decades. It has achieved to more than 50k published papers in each year . In order to find out the reason of this high interest to these materials, the first question is to be asked “What is the reason arousing a lot interest to these 2D materials by the scientific community?”. There are probably three answers that we could response for this question. The first answer lies within the unpredictable properties which is not seen any of ever tested materials. One example would be the observed ballistic transport throughout sub-micrometer distances in graphene in which no other film of similar thickness is known to be even poorly metallic or continuous under ambient conditions [1]. Second example would be change observed in band structure of 2D materials, as the thickness of the material reduced from bulk to monolayer [3]. The number of examples of these unpredictable properties can be increased up that it will be mentioned in this review article in further sections. The second answer is that the number of new 2D materials discovered which are  now exceed a hundred [4] and continues toincrease. A new material means variety of distinct research to explore the potential of that material by different scientific groups which results in increasing the number of publications. Thirdly, this 2D materials family includes all types of the materials that we know from the 3D bulky world such as semiconductors, metals, superconductors, insulators, semimetals and etc [4]. For these reasons,  2D layered materials are expected to be the basis of next generation  electronic circuitry.

Some of these recently studied 2D materials are plotted in **Fig.1** in terms of number of published papers over the years. The materials included into the plot are not the only materials in this category, however, they are probably the most studied ones.  This plot includes, graphene (semimetal), $MoS_2$, $WS_2$, $MoSe_2$, $WSe_2$, $MoTe_2$, $WTe_2$, $TiS_2$ and $SnS_2$ (semiconductor) so called transition metal dichalcogenides (TMDCs), elemental 2D materials such as phosphorene, silicene, germanene, borophene (semiconductor and metal), Mxene 2D materials (most of them metallic), $NbS_2$ (superconductor). The graphene is still number one in the list in terms of the number of the published papers compared to that of the other materials and enormous difference in total number of the published papers which have been still dominated by the graphene papers. On the other hand, it is seen from the plot that it is the earliest ever started to impact on the 2D research, which is around 2005 compared to the other materials. $MoS_2$ is probably the second most studied material within this 2D materials family. Although, its potential has been introduced in 2005 very similar to graphene [5], it is clearly seen that $MoS_2$ and all the other 2D materials almost have started to impact on around 2010.

It is interesting to see from the plot that the number of publications is saturated for the last few years for almost all the 2D materials except for the Mxene materials which is still showing significant increase. The saturation in the research field might indicate two important points; no extra possible data addition into the field and/or end of the laboratory directed research and developments. In both cases, beginning of a new phase has been starting for these materials that we will see for the next decade, which is called industrialization. At this point, one remarkable drawback has been seen for these materials, which is the limitation to obtain high quality and large area materials. Graphene and some other 2D materials such as hBN passed this stage successfully. Nowadays, roll to roll growth is possible for graphene on Cu foils with high quality and large area [6-8]. It is also possible to grow hBN in a cm scale by Chemical Vapor Deposition (CVD) [9]. However, it is still a big issue for many other 2D materials especially for the TMDCs and the elemental 2D materials such as phosphorene or borophene [10]. In this review article, corresponding literature about the methods of obtaining continues and large area 2D TMDC materials, especially sputtering and the device applications from those continues films will be discussed in detail.

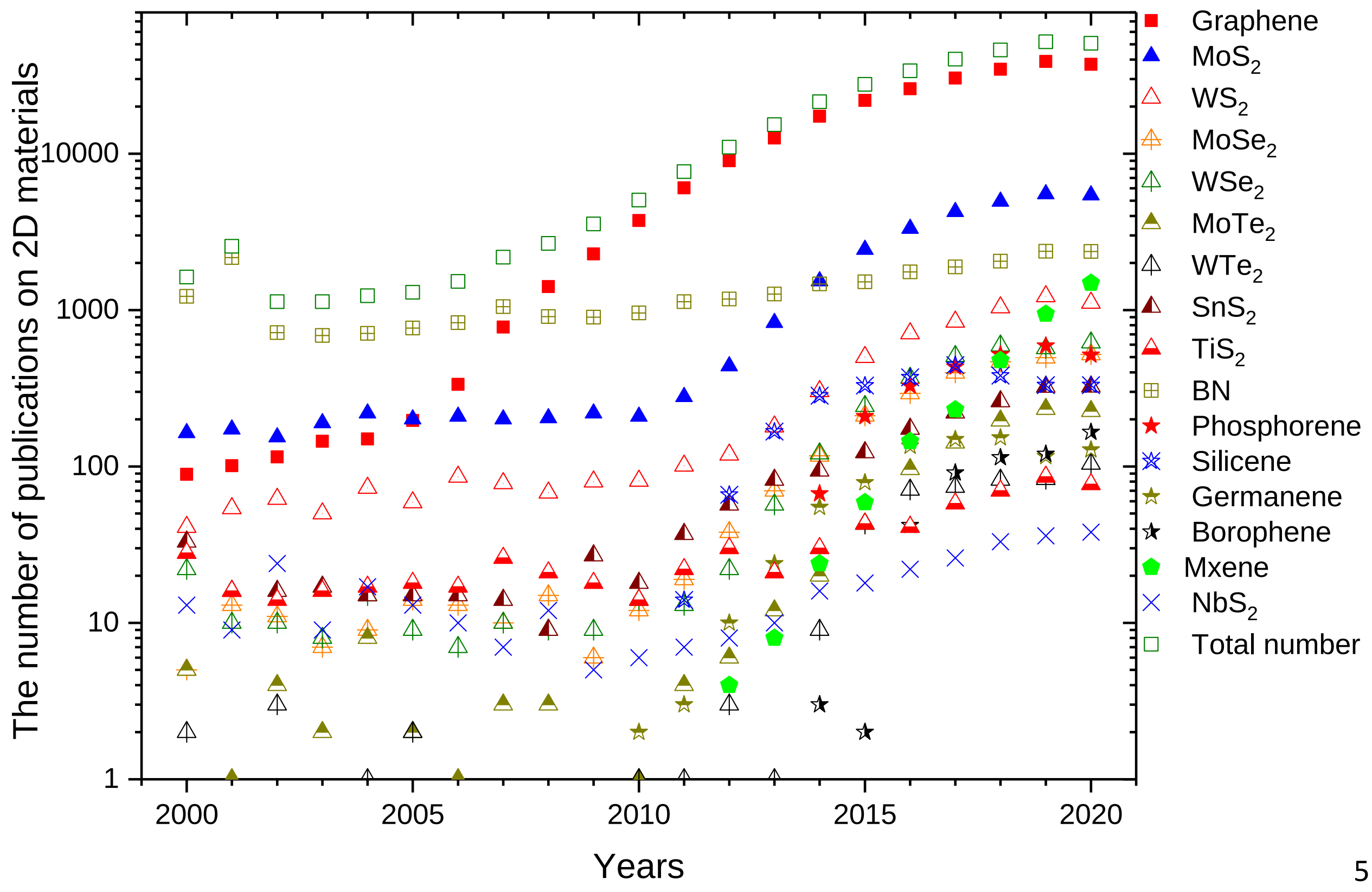

**Figure 1.** Published papers on the 2D materials over the years. “Triangular” symbols are for TMDCs, “stars” for elemental 2D materials, “squares” for graphene and total number of publication, “pentagon” for Mxene and “x” for the superconductor $NbS_2$.

## 2. Large area and continues 2D materials growth methods

There are, actually, many different available growth methods reported to obtain a high quality, large area TMDC materials, however, it is reported that the largest domain sized 2D materials are obtained by the mechanical exfoliation from the bulk materials [10]. Although, it is achieved very good quality single crystal materials with this method, it brings along a lotsof issues. First of all, repeatability is the big issue which affect the performances of the fabricated electronic devices every time. In addition, it is not suitable for mass production as well as not an efficient method which prevents it usage in industry. Lastly and probably the most important one is the uncontrollable interface contamination which has detrimental effects on the device performances [11]. Therefore, it is required to grow high quality TMDC 2D materials with other methods

The common bottom-up methods which are mostly employed to grow high quality TMDCs are CVD, metal organic CVD (MOCVD), molecular beam epitaxy (MBE), atomic layer deposition (ALD), pulsed laser deposition (PLD) and radio frequency magnetron sputtering (RFMS). CVD (powder vaporization) is most commonly used method among all. Although, it is possible to obtain high quality crystals in this method, it is not suitable for industrial processes compared to the ALD, RFMS, MOCVD and PLD methods due to the very low growth rates [11]. On the other hand, CVD method does not have very good control on deposition properties such as vapor rates of the constituent powders and often results in a gradient of TMDC properties even on the same wafer. These fundamental properties of the powder vaporization CVD methods limits to grow large area materials in a single domain. However, it might become possible in future to grow larger 2D TMDC materials with the help of new growth strategies developed in CVD such as used promoters and etc [12, 13]. These kinds of developments in CVD method make it very promising method for growing large area TMDCs in near future.

Among mentioned methods, RFMS and MOCVD growth methods are probably the only methods which are currently and effectively used in industry. Especially, MOCVD holds its position on all of the epitaxial growth of LEDs and laser diodes in all colors (blue, green, red) [14-16] and high power infrared laser diode industry [17]. On the other hand, RFMS growth methods is also leading many important coating industries. In the next section, a detailed information about the RFMS method is included.

### 2.1. RFMS growth method

Once the planar magnetron sputtering was introduced by the J.S. Chapin in 1974 [18], RFMS became one of the most desired coating and evaporation system especially preferred in industry. The threshold in this achievement was to obtain high deposition rates. Nowadays, it is possible to achieve 1nm/s to 10 nm/s for various of depositions by RFMS [19]. RFMS method has been used for hard, wear-resistant coatings, low friction coatings, corrosion-resistant coatings, decorative coatings and coatings with special optical, or electrical properties in industry more than for almost fifty years [20]. The main advantage of the sputtering method among the other coating methods, which makes it preferable in industry, is the reliability, large area homogenous growth and the repeatability of the grown thin films. The method has a good control on the grown thin films' microstructures, surface morphologies, optical and electrical qualities even the chemical compositions [21]. Growth conditions controlled by growth pressure, base pressure, RF or DC power, substrate temperature and etc. might lead to obtain a thin film from an amorphous film phase to high degree crystalline film or insulator to highly conductive film. This brings about the growth optimization in order to achieve thin films with desired characteristics. The main growth parameters controlled by these growth conditions are the growth rates, the mean free path of the sputtered ions from the target and their energies. If the growth pressure is relatively high, it results in lower growth rates which generally requires for obtaining high optical and smooth surface quality thin films. However, achieving high conductivity in metal films requires growth at low pressures and high RF or DC powers [22]. Therefore, growth optimization for thin films to be grown is one of the most important steps in order to obtain right quality thin films.

The sputtering method relies on the bombardment of target material via the ions created by the plasma produced with DC and RF gas discharge. **Fig. 2** shows the schematic diagram for sputtering. Plasma is mostly produced from the Ar gas, since it is easy to obtain at high purity with a low cost compared to the other noble gases such as Kr and Xe. It is also more effective in terms of the sputtering yield, which is the ejections of ions from the target per ions hitting the target, than that of the He and Ne gases. The color of the plasma produced by the Ar gas is blue, although it is most of the time seen in pink color due to the oxygen found in the plasma. Therefore, relatively low base pressure plays an important role to remove the oxygen from the chamber during deposition. The growth process generally performed at which the pressure is varying from 1mTorr to100 mTorr. The mean free path of the particles varies from a few mm to cm within this pressure range. The energetic ions hitting the target material cause ejection of ions from the target material by the transferring momentum from Ar ions in the plasma which causes to erosion of the target surface. Those ions sputtered will be collected on the substrate which is suited geometrically in different positions. The position of the substrate and the target might have also impact on the grown thin film properties. The higher the RF or DC power applied on the target, the higher the energies of the particles ejected from the target. It is good to note that energies of the ejected ions from the target depends on the momentum transferred from the ions in the plasma. This brings the importance of the target density. The higher the density, the larger the sputtering yield from the target [19]. Sometimes this high energy of the ejected ions might be an issue due to the creation of defects on the substrate.

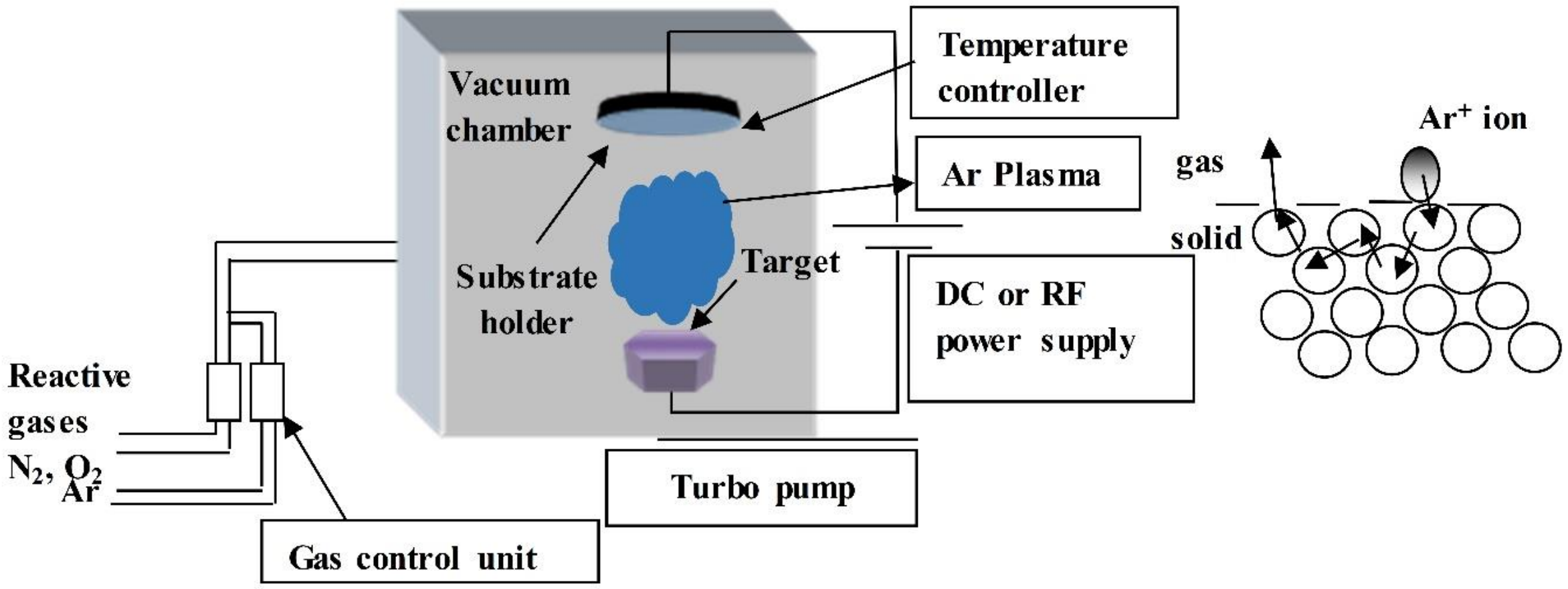

**Figure 2.** Schematic diagram of the sputtering method. The left panel shows the common equipment for the sputtering. The right panel shows the ion collisions. The figure is reproduced from the permission [23].

Variety of materials can be grown by RFMS especially for semiconductor thin films, nitride, carbide, oxide or chalcogenide films. Although it is not very easy to find target materials for any oxides, nitrides or chalcogenides, it is possible to grow reactively any oxide or nitride materials from the elemental targets by just inserting a related gas into growth chamber. It might be even possible to grow reactively chalcogenides by using $H_2S$ [24] or $H_2Se$ [25] gases. In reactive sputtering, oxygen or nitrogen gas can be supplied into the chamber before the growth process. The quantity of the reactive gas is very important to control the compound thin film' properties. For example, it is common to accept that if the reactive oxygen partial pressure percentage to the total growth pressure is relatively high, perhaps the compound films might become amorphous. On the other hand, with the ratio about 10% or lower, it might be highly possible to grow high quality oxide thin films [26].

## 3. Growth of 2D TMDCs with RFMS

Although RFMS growth method is one of the most desired system in industry for variety of thin film growths, it has a deficiency to grow high quality in terms of their structural and optical characteristics of 2D TMDC materials. The deficiency is its limited control on the stoichiometry of the TMDCs due to very high vapor pressure of the chalcogenide species compared to the metal in the TMDCs. This basically causes two different phenomena during the sputtering process. First one is the well-known preferential sputtering in which the sputtering from a TMDC target yields different sputtering rates for the metal and chalcogen species in the target, which is apparently larger for chalcogen atom [27]. This give rise to always larger content of the metal left on the target surface depending on the sputtering time which might be the one aspect to obtain non-stoichiometric films. However, this may not seem the case causing the chalcogen deficient films. Since the sputtering yields are not same for chalcogen and metal species, it is likely possible that the amount of chalcogen and metal atom sputtered from the target might be equal even chalcogen amount is lower than the metal atom on the surface of the target due to the difference in the sputtering yields. This might equalize the fluxes of the species in the plasma [28]. Second phenomenon is the desorption at the substrate which is more in the chalcogen than the metal species during the growth. In order to obtain higher quality 2D TMDCs, controlling the substrate temperature is one of common applied parameters during the sputtering growth. Therefore, it is indispensable to obtain a chalcogen deficient films in the sputtered TMDC layers. However, a good strategy for RF sputtering of $MoS_2$ layer growth has been applied to obtain better films with simply; room temperature and low temperature growth followed by a post annealing at high temperature [29]. It has shown that the conductivity of the 3-4 layers of RF sputtered $MoS_2$ layers' conductivity is comparable with the layers produced by the most attractive methods CVD and mechanical exfoliated layers. It is also good to note that stoichiometry is more important issue for the ***S***-element based TMDCs compared to that of ***Se*** based TMDC materials due to the much lower vapor pressure of the ***Se*** element than ***S***. There have been so far non-stoichiometric TMDC 2D layers reported already grown by sputtering as well as for variety of the different methods [30-33].

There are many applied growth strategies to grow 2D TMDC layers by sputtering method. A most common and easiest one is the direct sputtering from TMDC targets of $ReS_2$, $MoS_2$, $WS_2$ or $MoSe_2$ [32-37]. Although it has shown a good control on many important parameters, such as crystal structure, surface morphology, growth mode and etc., of 2D $WS_2$ layers with one step, direct sputtering by our group, 2D $WS_2$ layers are not stoichiometric as discussed above. In order to solve the stoichiometry problem as well as to improve the structural quality in sputtered TMDCs 2D materials, very successful growth strategies have been suggested to date. It is possible to classify these strategies into two parts; single (or one) step and two step growths. For the one step growth method, the easiest is to grow by the reactive sputtering with $H_2S$ gas from a TMDC target [24]. For example, $MoS_2$ target is used for reactive sputtering in Ar/$H_2S$ mixed gas environment and it has shown that an improved crystal and stoichiometric properties is obtained in sputtered 2D layers in mix Ar/$H_2S$ gas environment. Another successful strategy has been applied by using $H_2S$ plasma assisted synthesis of either sputtered Mo and thermally evaporated $MoO_3$ layers [38]. It has been concluded that very uniform high quality and large area $MoS_2$ few layers are possible to grow with this $H_2S$ plasma assisted method. Similarly, Mo metal target sputtered in vaporized sulfur ambient has shown very successful wafer scale high quality $MoS_2$ 2D layers [39-41]. One example is given below in **Fig. 3** which shows microscopic and AFM images of a large area continuous thin films obtained by sputtering under sulfur vapor. A "cm" scale $MoS_2$ layer has been clearly achieved successfully. In addition to these, thickness control from monolayer to a few layer has been indicated with the Raman shift in $MoS_2$ vibrational modes [39]. Similar thickness control in a "cm" scale 2D material is also achieved with a similar method [40]. The quality of the layers obtained with the mentioned method is comparable with TMDC layers obtained by the most commonly preferred methods such as CVD and mechanical exfoliation. On the other hand, it is possible to have large area uniform continuous layers indicating the methods' advantage over the mentioned growth methods. It is also keep in mind that one step growth is always preferable compared to the two step processes, since it is easy and suitable for industrial applications.

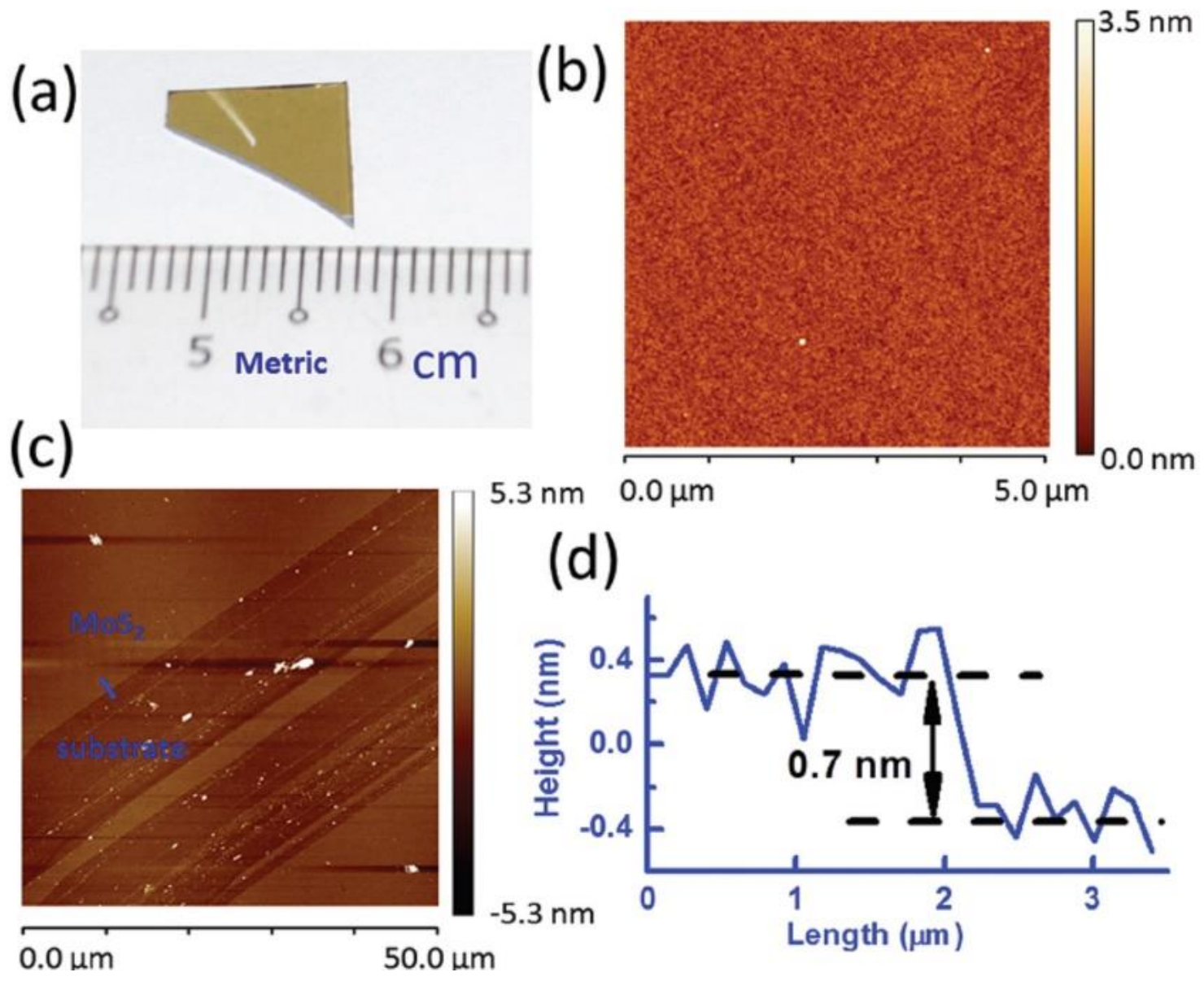

**Figure 3.** One step sputtered characteristics of the $MoS_2$ layer. Adapted with permission from ref. [39]. Copyright (2021) American Chemical Society.

As for the two step growth method, the strategies employed in the literature are as follows;

- Sputtering TM (Mo, W, Nb) layers + CVD post-sulfurization or selenization [42-47]
- Sputtering of the $MoO_3$ or $WO_3$ layers + CVD post-sulfurization [48-50]
- Sputtering $MoS_2$ + CVD post-sulfurization [51, 52]
- Sputtering $MoS_2$ or $WS_2$ + e-beam irradiation [53]
- $MoS_2$ growth + laser annealing [54]

The most commonly employed method from the two-step growth strategies given above is the sulfurization and selenization of the Mo and W sputtered layers in order to obtain large area 2D TMDC layers. It becomes even possible to obtain $WS_2/MoS_2$ hetero-structures with sulfurization of the sequentially sputtered Mo and W metal seed layers [45]. One of the earliest published study on two-step process has shown the thickness dependence of the Mo or W sputtered layers effects on the ultimate product material of $MoS_2$ and $WS_2$. The formation of a large-area, textured few layers MoS2 and WS2 films with polycrystalline nature employing this method have been indicated. Furthermore, it has also concluded that the thickness of the metal seed layers of Mo and W is critical to observe a horizontal growth or to a vertical growth [44]. The sulfur amount has also been investigated on the pre-deposited Mo layer to form 2D $MoS_2$ layers. Uniform $MoS_2$ films have been obtained successfully in a large-area. It has been found that sulfur amount is not effective in controlling the numbers of layers. However, small Mo oxide clusters at the surface of the thin film have been indicated  the sulfur deficient growths by TEM images [42]. Similarly, 2D wafer scale $MoS_2$ obtained by employing a $MoO_3$ sputtering and sulfurization process [48, 49]. **Fig. 4** shows a 4-inch wafer scale $MoO_3$ converted into the 2D $MoS_2$ layers on a sapphire substrate. Homogeneous layer formation has been tested by the Raman and PL mapping. The Raman studieshave resulted in 24.9 $cm^{-1}$ wavenumber difference between the $E_{2g}$ and $A_{1g}$ vibrations confirming almost four layers of $MoS_2$ formation. Raman and PL mapping has shown highly uniform layers on the entire 4-inch sapphire substrate [48]. Although there only a few studies performed to produce high quality 2D TMDC layers by sputtering from TMDC target ($MoS_2$ and $WS_2$) and following sulfurization two-step process [51, 52], the process has also shown the formation of a possible large area with high quality TMDC 2D layers. A good crystalline $MoS_2$ formation after sulfurization as well as the enhanced stoichiometry have been revealed [52].

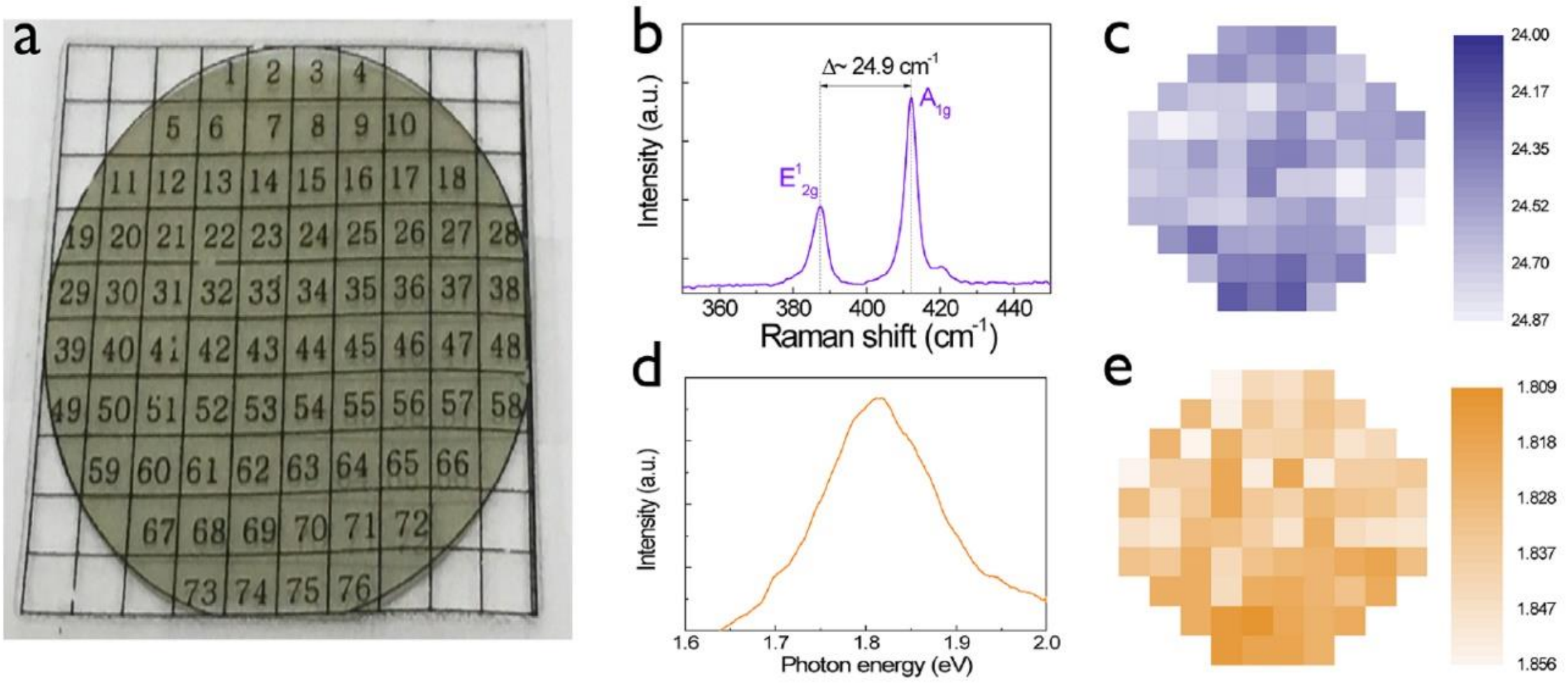

**Figure 4.** Wafer scale integration of the $MoS_2$ from sputtered $MoO_3$. Adapted with permission from ref. [48]. Copyright (2021) American Chemical Society.

The growth mechanism of the sputtered most TMDC materials includes two stages: the first stage is layer by layer growth through basal plane (002) parallel to the surface of the substrate and the second stage is vertical growth perpendicular to the substrate that is to the (100) plane. The first stage is well known 2D growth mode which is seen in many different bulk materials such as GaN or ZnO [55, 56]. The second stage is the 3D growth mode in which island like nucleation occurs and the vertical growth becomes favorable which is also seen in many different materials [57, 58]. Anyone can see these stages in any sputtered TMDC materials, if the thickness of the materials is large enough. This growth mode actually is well-known Stranski-Krastanov growth mode in which the material starts to grow layer by layer 2D mode for a certain thickness and then continues with the 3D vertical growth on top of that 2D layers [32]. 2D layer thickness might be changed and controlled with some sputtering parameters and it might be between 2-50 layers [28] which corresponds to roughly a 30 nm thickness for the $MoS_2$ or $WS_2$. However, this claim needs more detailed investigation in order to understand clearly about 2D layer thicknesses in different growth conditions. On the other hand, it has been shown that the highest (002)/(100) orientation ratio is obtained with growth rates near 1 atomic layer per second with high ion energies in the plasma. Whereas for growth rates of less than one monolayer per second coupled with low ion energies and fluxes, growth of $MoS_2$ has shown a strong (100) orientation [59]. We previously have also shown that the low growth pressure of $WS_2$ has given rise to a columnar growth which is not possible to obtain 2D materials [32]. The figure from our previous study given below summarizes the concept given above. If you look at the SEM images given in the figure, only a smooth surface morphology can be seen in samples grown at 5 and 10 s as shown in **Fig. 5 (c)** and **(d)**. As the growth time gets longer, 3D structures emerge first in the $WS_2$ thin film grown for 20 s duration as bright points in the SEM images (**Fig. 5 e).** The number of these 3D structures increases for further increase in the growth time, as shown in **Fig. 4 (e)–(g)**. When the XRD data and SEM images are considered for the thin films grown at 5 and 10 s duration, it is noticeable that there is no diffraction peak other than (002) that corresponds to the layer by layer growth parallel to the substrate. Therefore, it might be speculated that first a few layers of $WS_2$ grow on the substrate layer by layer, basically 2D growth, and then 3D structures, 3D growth, start to appear from some nucleation layers [32].

This Stranski-Krastanov growth mode in sputtered TMDC materials has been realized in many other different groups. One of the very early transition from 2D to 3D growth mode in $MoS_2$ was proved with TEM measurements in 1992 [60]. They clearly showed the vertical formation of the $MoS_2$ layer on a continuous few layer thick 2D $MoS_2$ layer. Similarly, 2D layer by layer growth transition to 3D growth mode has been realized in even CVD grown $MoS_2$ flakes [61]. Furthermore, it has been shown that sputtered metal seed layer of Mo or W thickness is effective to control the 2D growth or 3D growth such that as the Mo seed layer thickness decreases, a morphological transition from vertical-to-horizontally grown $MoS_2$ layers becomes more pronounced [44]. Sulfur content effect on the two-step grown layers also has been investigated [42]. It has been speculated that there is a competition between the Mo oxide segregation to form small clusters and the sulfurization reaction to form planar $MoS_2$ film which depends on the amount of sulfur content. There are several mechanisms explained for the formation of the 3D growth stages. One of them is the increasing defect and strain energy related after certain thickness causes the nucleation point in which the 3D structures forms [61]. On the other hand, low surface energy of the (002) basal plane disruption with the adsorption of contaminant oxygen from ambient water vapor is reported to be another reason [60]. In addition, it has been reported that the volume expansion plays an important role when conversion from metal layer to the TMDC occurs depending on the metal layer thickness [44]. All these explanations clearly demonstrate that it is possible to obtain 2D TMDC structures, which is very important both for transport devices such as field effect transistors (FETs) or the optoelectronic devices such as photodetectors. The following chapters will be dealing with those device applications to date on the sputtered TMDC layers.

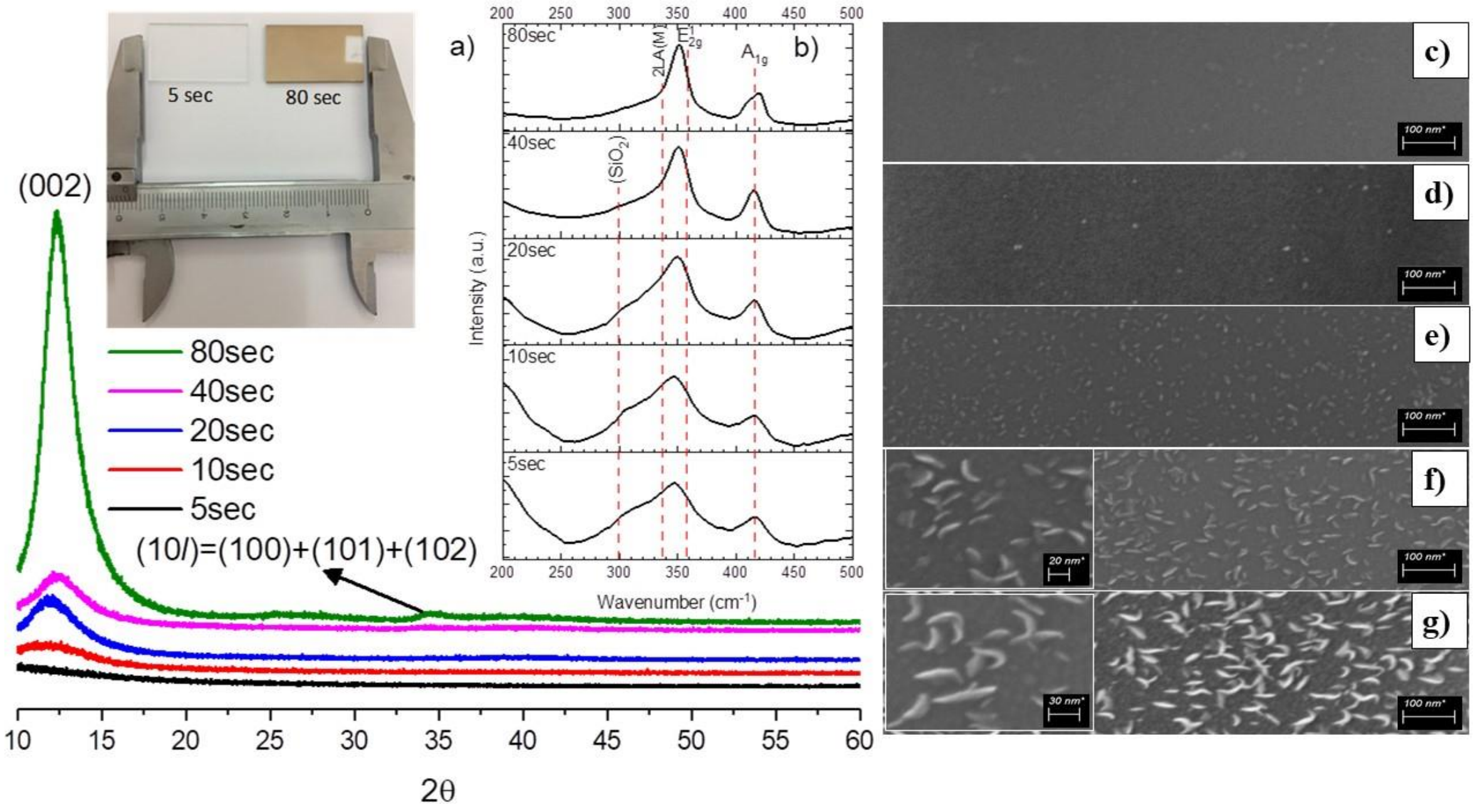

**Figure 5. (a)** XRD and **(b)** Raman measurement results of $WS_2$ films grown on the glass substrate. SEM images at **(c)** 5 s, **(d)** 10 s, **(e)** 20 s, **(f)** 40 s, and **(g)** 80 s. Inset shows pictures of large area growth of 2.5 cm × 1.0 cm size of continuous $WS_2$ thin films on the glass substrate. The sample grown in 5 s duration is highly transparent, while the sample grown in 80 s is brownish Adapted with permission from ref. [32] Copyright (2021) American Chemical Society.

## 4. 2D sputtered TMDC device applications; FETs and photo-detectors

Since the size of the high quality TMDC layers grown is limited, this brings many problems with itself especially for the fabrication of the solid state devices. The use of e-beam lithography makes whole device fabrication process so complicated as well as bringing the repeatability issues. On the other hand, sputtered TMDC layers makes fabrication of solid state devices very easy compared to that of CVD grown or the exfoliated TMDC layers with a single step photolithography process. However, sparse reports have been realized so far on the devices fabricated with the sputtered TMDCs. One needs to keep in mind that most of these sputtered TMDC layers have grain boundaries smaller than 1 μm, which might be the limit of the method to produce materials as high quality as layers produced with CVD and exfoliation. However, it has been achieved very promising high quality device characteristics so far. We will review the device characteristics in the next section.

### 4.1. FET Applications of sputtered TMDCs

FETs are voltage controlled with large input impedance devices serving as one of the main components for modern electronic circuits. It has been utilized for many different applications such as high power and frequency switching applications, RF amplifiers, voltage controlled resistors, logic circuits etc. [62]. Scaling down the Si-based FETs have revealed so many problems such as loss of channel controllability and the short channel effect, which prevents the reduction of further transistor sizes [63]. Although, there are many different alternative solutions to these problems such as the FinFETs, 2D materials holds a position to be the next alternative materials. 2D TMDC semiconductors have been employed as the channel materials due to their important material properties such as having a naturally passivated surfaces with no dangling bonds, tunable properties such as bandgaps, having a high mobility (at least in theoretical limits) [64, 65].

**Table 1.** Comparison of the fundamental characteristics of the various sputtered TMDC based FETs.

| Structure | Method | Measurement condition | Mobiliy ($cm^2$/V.s) | $I_{on}/I_{off}$ ratio | Ref. |
|---|---|---|---|---|---|
| $MoS_2$/PDMS* | Sputter + annealing | Vd=1V | 0.024 | $10^2$ | [66] |
| $MoS_2/SiO_2$/Si | Sputter | Vd=10V | 24.17 | $10^6$ | [67] |
| $MoS_2/SiO_2$/Si | Sputter + CVD | Vd=5V | 0.8 ± 0.2 | Log 4.3 ± 0.7 | [48] |
| $MoS_2/SiO_2$/Si | Sputter + CVD | Vd=0.5V | 12.24 | 1.57 × $10^6$ | [68] |
| $MoS_2/SiO_2$/Si | Sputter | Vd=4V | ~12.2 (hole) | $10^3$ | [39] |
| $MoS_2/SiO_2$/Si | Sputter + CVD | Vd=1V | ~21 (bilayer) ~25 (trilayer) | ~$10^7$ (bilayer) $10^4$-$10^5$ (trilayer) | [49] |
| $MoS_2/SiO_2$/Si | Sputter + CVD | Vd=1V | ~29 (~1.4 nm) max. 181 (~3.8~6 nm) | $10^5$ | [69] |
| $MoS_2/SiO_2$/Si* | Sputter $MoS_2$ + CVD | - | 0.0136 (bilayer) 0.0564 (five-layer) | $10^4$ | [51] |
| $MoS_2/Al_2O_3$/MgO* | Sputter + CVD | - | 0.005-0.01 | - | [70] |
| $WS_2/SiO_2$/Si | Sputter+CVD | Vd=1V | 17-38 | $10^4$-$10^5$ | [71] |
| $WS_2/SiO_2$/Si | Single-Step sputter | Vd=10V | 10-20 | ~$10^4$ | [72] |
| $MoSe_2/WSe_2/SiO_2/Si$ | Sputter +CVD | Vd=1V | 2.2(hole)<br>15.1(electron) | ~$3.6 \times 10^5$ | [73] |

*Top gated device configuration

Many FETs have been investigated so far in top-gate or back-gate configurations by using sputtered $WS_2$, $MoS_2$ or $MoSe_2/WSe_2$ heterojunction channel materials in which the details are summarized in **Table 1**. Single-step sputtering from TMDC targets or sputtering under the sulfur vapor environment, or two-step sputter + sulfurization/selenization of the metal seeds of TMDC materials have been applied for the FET materials as also mentioned in the growth of TMDC section. One of the best FETs obtained on the sputtered $MoS_2$ after post-deposition annealing at 700 ° C in an environment of sulfur and argon layer has achieved the value of 181 $cm^2$/V.s mobility which is comparable among the TMDC FETs produced by any other methods [67]. **Fig. 6 (a)** shows this two-step process to obtain high quality $MoS_2$ 2D layer, while **Fig 6 (b)** shows the output and transfer characteristics of fabricated FET. Transfer characteristic of the fabricated device has shown an almost 7 orders of $I_{on}/I_{off}$ ratio which is also quite high. The high mobility value and remarkable characteristics of the device fabricated are claimed due to the improved crystal quality of the $MoS_2$ layer with the two-step methods.

Our previous study shows one of the single step, large area, continuous $WS_2$ films grown on $SiO_2$ substrate FET device, as illustrated in **Fig. 6 (c)-(e)** [71]. **Fig. 6 (c)** clearly shows the contrast between variable thickness grown $WS_2$ layer and the $SiO_2$. As can be seen from the **Fig 6 (d)**, the transistor has exhibited ambipolar behavior and approximately 20 $cm^2$/Vs mobility with an almost$10^5$ $I_{on}/I_{off}$ ratio. **Fig. 6 (d)** displays the transfer characteristics of the $WS_2$ FET device on the 1 s grown continuous film. It can be observed dominant p-type channel behavior in the transfer characteristics with the increase in the current at the negative voltage side more than the positive side, as the positive voltage is applied between the drain and source. In addition, the sputtering method has been used to grow TMDC heterojunction transistors. In this case, the sputtering method is easiest to produce vertical heterojunctions. A vertical $MoSe_2/WSe_2$ p–n heterostructure using a sputtering-CVD method has been reported [72]. The heterostructure is formed as follow; first W film was sputtered on $SiO_2$/Si substrate by using sputter, and then sputtered W film was further put in a furnace for selenization. After the selenization process of the $WSe_2$, a Mo film was sputtered on it. The second selenization process was helped to form $MoSe_2/WSe_2$ heterojunction. $WSe_2$ and $MoSe_2$ back-gate field-effect transistors have exhibited good gate modulation behavior and high-hole 2.2 and 5.1 $cm^2$/Vs electron mobility values, respectively. As can be seen from the **Table 1**, acceptable device performances have been reported. It needs to be pointed out that the device performances are not as good as the devices produced by the CVD method or the exfoliation method. The common mobility value achieved on these sputtered TMDCs FETs varies between 10-20 $cm^2$/V.s while $I_{on}/I_{off}$ ratio varies between $10^2$-$10^7$ [39, 48, 49, 51, 66-72].

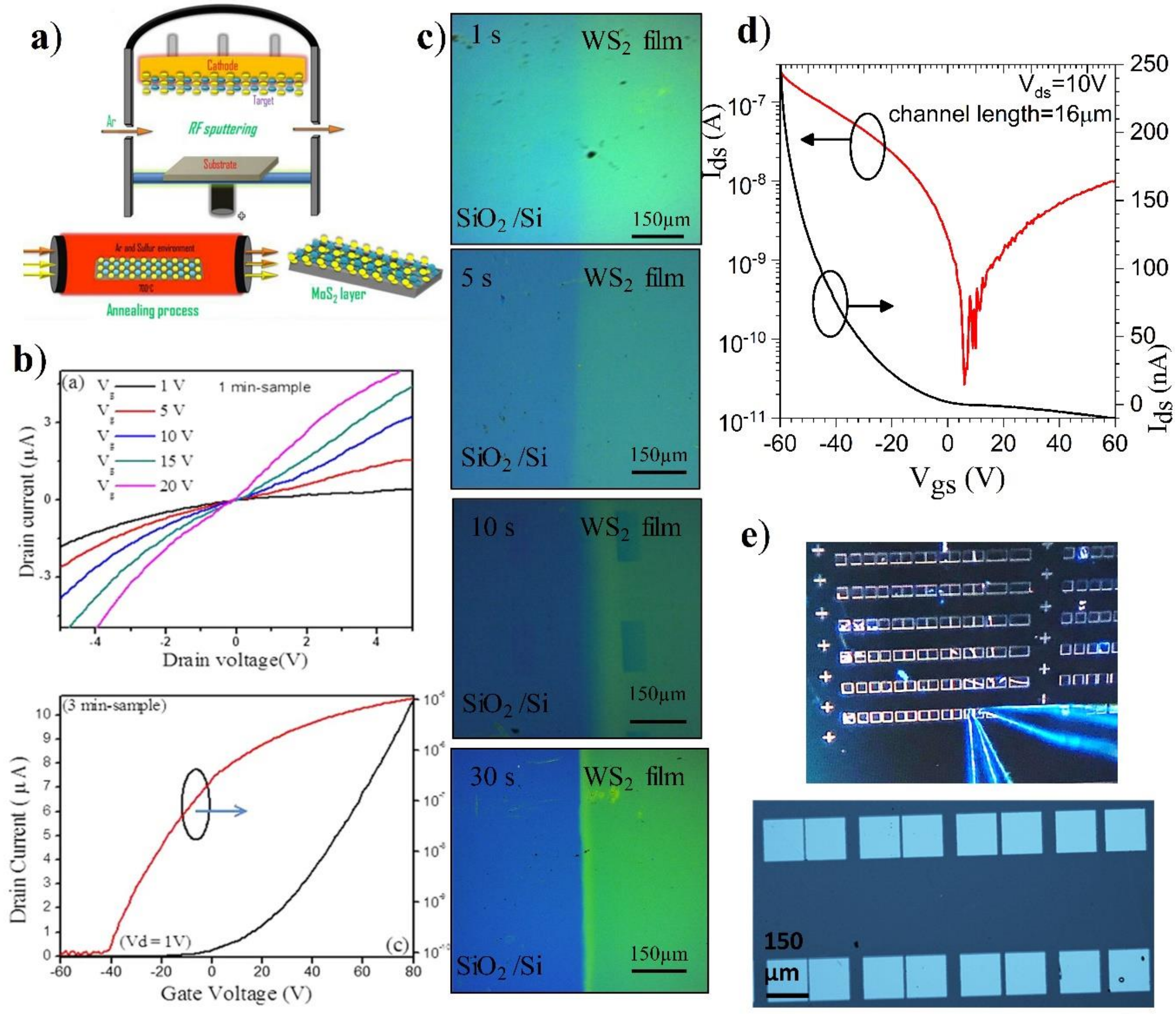

**Figure 6. a)** Schematic representation of the two-step system to prepare continuous TMDC films. **b)** $I_d$–$V_d$ and $I_d$-$V_g$ of $MoS_2$ FET adapted with permission from ref. [69]. **c)** Optical Microscope image of the continuous single-step sputtered $WS_2$ films grown variable thickness on Si/$SiO_2$ substrates **d)** $WS_2$ FET $V_{gs}$-$I_{ds}$ characteristic **e)** Dozens of $WS_2$ FET fabricated on same substrate. Adapted from ref. [72].

### 4.2. Photodetector Applications of sputtered TMDCs

Photodetectors are one of the main application areas for the 2D TMDC materials in the visible regions of the electromagnetic spectrum because of some properties such as the large absorption coefficient of the TMDC materials. It is shown that a few layers of $MoS_2$ absorbs 5-10% of the sunlight which is quite good. The absorption coefficient of the TMDC materials is about $10^5$-$10^6$ $cm^{-1}$ in the visible range [74, 75]. Furthermore, ultrafast response speeds in photodetectors have been achieved in these materials [76]. Therefore, intensive efforts have been carried out on the TMDC photodetectors. On the other hand, there are sparce number of studies performed in order to reveal the potential for the photodetectors of the sputtered TMDC layers. **Table 2** shows the characteristics of the photodetectors fabricated on the sputtered TMDC materials.

One of the early sputtered TMDC photodetector has shown rather good performance even compared to that of two-step processed (wet chemical process + sulfurization in CVD), CVD grown or the mechanically exfoliated high quality $MoS_2$ layers [76]. $MoS_2/S_i$ heterojunction has been produced by the direct sputtering of the $MoS_2$ on the Si substrate followed by the annealing at 800 °C in Ar atmosphere. Remarkable value of detectivity of almost $10^{13}$ Jones has been achieved indicating a good potential of the sputtered TMDC layers to be an photo-detector. The direct sputtering method at low temperatures makes TMDC materials suitable for flexible photodetectors. Recently, it has been shown that $WS_2$ sputtered on spin coated polyimide substrate followed by the e-beam irradiation has shown a maximum of $10^7$ Jones responsivity. It can be concluded from the table that the most of the sputtered TMDC photodetectors shows a fall and rise time within a few μs range with larger fall time which is showing a recovery issue of the photodetectors. On the other hand, appreciable responsivity values of around a few A/W together with the larger than $10^8$ Jones are also quite promising for the future of these sputtered TMDC materials in many different applications.

**Table 2.** Comparison of photo-response characteristics of present sputtered TMDC based photodetectors.

| Structure | Measurement condition (wavelength, bias) | Rise time (s) | Fall time (s) | Responsivity (A/W) | Detectivity (Jones) | Ref. |
|---|---|---|---|---|---|---|
| $MoS_2$/p-Si | V=0V | 38.78 µs | 43.07 µs | 0.03 | $1.45 \times 10^{10}$ | [77] |
| Pd-$MoS_2$/n-Si | V=0V | 2.1 µs | 178 µs | 0.654 | $1.0 \times 10^{14}$ | [78] |
| $MoS_2$/p-Si | V=2V | 10 µs | 19 µs | 8.75 | $1.4 \times 10^{12}$ | [79] |
| a-$MoS_2$/$SiO_2$/Si | 473-2712 nm | 10 ms | 16 ms | 47.5 | $1.26 \times 10^{7}$ | [80] |
| $MoS_2$/p-Si | V=0V 808 nm | 3 µs | 40 µs | ≈0.3 | ≈$10^{13}$ | [76] |
| $MoS_2$/p-Si | 400-850 nm | - | - | - | - | [24] |
| $WS_2$/$SiO_2$/Si | 365 nm | - | - | 53.3 | $1.22\times 10^{11}$ | [81] |
| $WS_2$/$SiO_2$/Si | 450-635 nm | - | - | $0.19\text{-}2.45\times 10^{-3}$ | $1.25 \times 10^{7}$ | [82] |
| $MoS_2$/n-Si | 400-1200 nm | - | - | 1.8 | $5\text{x}10^{8}$ | [40] |
| $WS_2$/Polyimide | 450 nm | 0.48-0.82 | 0.70-0.88 | 1.25x$10^{-3}$ | 2.52×$10^{7}$ | [83] |
| (EBI-treated) | 532nm | | | 1.66 x$10^{-3}$ | 3.34×$10^{7}$ | |
| | 635nm | | | 0.53 x$10^{-3}$ | 1.08×$10^{7}$ | |

## 5. Conclusion

2D TMDCs in the semiconductor group have attracted the most attention within the scientific community due to their important properties since graphene has been discovered. The need to have a high quality material for solid state devices requires a good control on the grown TMDCs during the growth. It has been indicated that sputtering is an excellent method to obtain TMDC 2D thin layers. It has a good control on the thickness of the material as well as the crystal quality. In order to improve even the quality of the sputtered layers, different growth strategies have been employed. Stoichiometric good crystal quality materials have been achieved compared to the most popular methods such as CVD and exfoliation. On the other hand, devices like FETs and the photodetectors fabricated on the sputtered TMDC layers have shown appreciable performances showing the potential of the sputtering method.